\documentclass[a4paper,10pt]{article}
\usepackage[utf8x]{inputenc}
\usepackage{graphicx}
\usepackage{amsmath}
\usepackage{cite}
\usepackage{authblk}
\usepackage{color}
\usepackage{url}
\usepackage{enumitem}

\usepackage[margin=3cm]{geometry}

\title{\bf Climate network and complexity based ENSO forecast for 2026}

\date{}

\author[1] {Josef Ludescher}
\author[2] {Jun Meng}
\author[3] {Jingfang Fan}
\author[4] {Armin Bunde}
\author[5] {Hans Joachim Schellnhuber}

\tiny
\affil[1] {Potsdam Institute for Climate Impact Research (PIK), Member of the Leibniz Association, 14412 Potsdam, Germany}
\affil[2] {State Key Laboratory of Earth System Numerical Modeling and Application, Institute of Atmospheric Physics, Chinese Academy of Sciences, Beijing, 100029, China}
\affil[3] {School of Systems Science, Beijing Normal University, 100875 Beijing, China}
\affil[4] {Institute for Theoretical Physics, Justus Liebig University Giessen, 35392 Giessen, Germany}
\affil[5] {International Institute for Applied Systems Analysis, Laxenburg 2361, Austria}

\begin{document}

\maketitle

\begin{abstract}
The El Ni\~no Southern Oscillation (ENSO) is the dominant driver of interannual global climate variability and can lead to extreme weather events such as droughts or flooding. Recently, we have developed several statistical approaches for early ENSO forecasting, in particular, its El Ni\~no phase. The climate network-based approach allows forecasting the onset of an El Ni\~no event or its absence about 1 year ahead \cite{Ludescher2013}. The complexity-based approach allows additionally to forecast the magnitude of an upcoming El Ni\~no event in the calendar year before the onset \cite{Meng2019}. Additionally, we have developed methods for forecasting the type (Eastern Pacific or Central Pacific) of an El Ni\~no \cite{Ludescher2023b} and for probabilistic forecasting of La Ni\~na and neutral events \cite{Ludescher2025}, also by the end of the calendar year before the event. Here we present the forecasts of these methods for 2026. The climate network and the complexity-based approach do not provide concurring signals for this year. The combined forecast indicates that a neutral event is more likely than an El Ni\~no. If an El Ni\~no develops in 2026, the complexity-based approach predicts a weaker event with a magnitude of $0.84\pm0.36$°C.
\end{abstract}

\section{Introduction}

\begin{figure}[]
\begin{center}
\includegraphics[width=8cm]{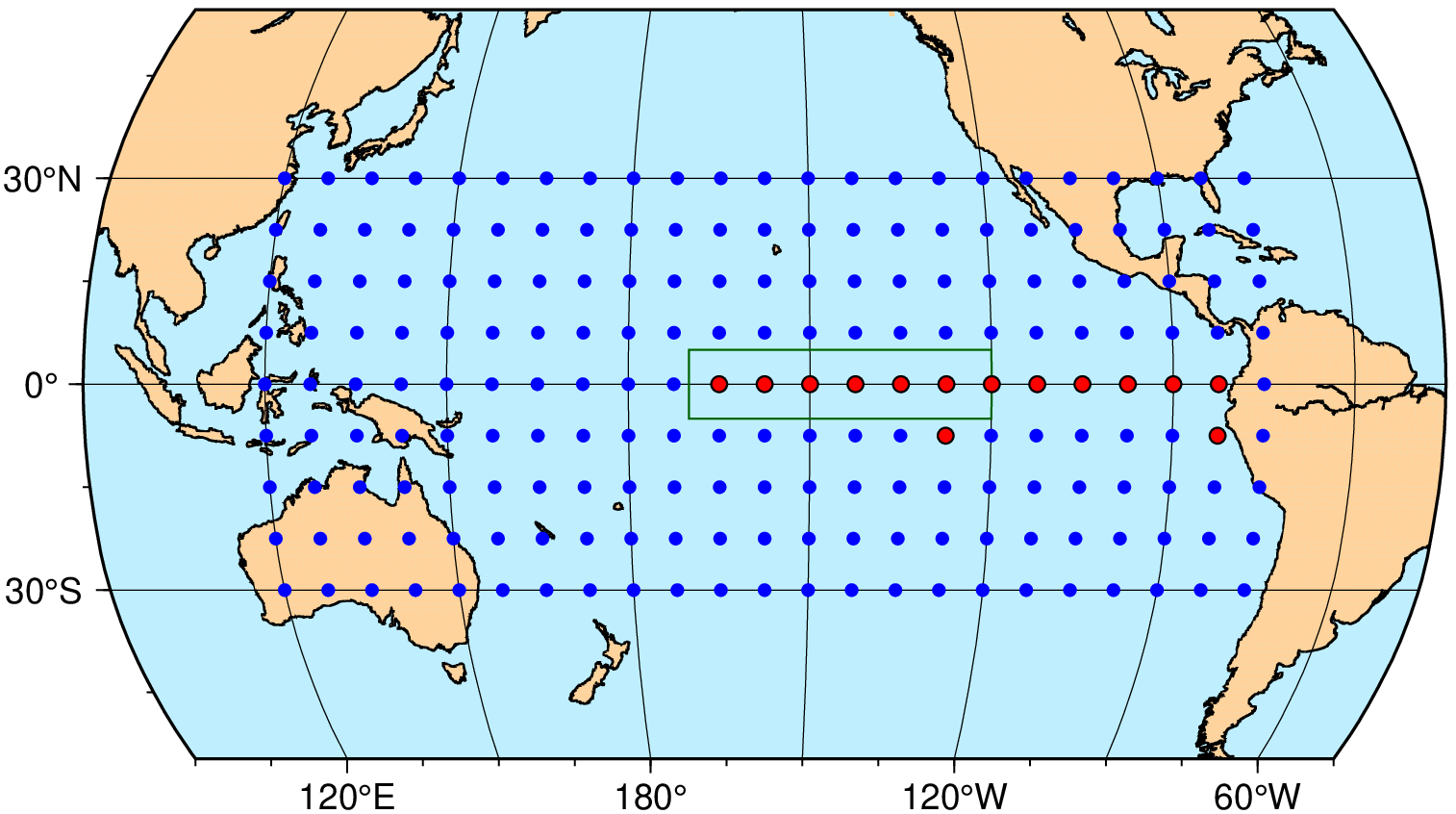}
\caption{{\bf The nodes of the climate network.} The network consists of 14 grid points in the central and eastern equatorial Pacific (red dots) and 193 grid points outside this area (blue dots). The green rectangle indicates the Ni\~no3.4 area. The grid points represent the nodes of the climate network that we use here to forecast the {\it onset} or absence of an El~Ni\~no event. Each red node is linked to each blue node. The nodes are characterized by their surface air temperature (SAT), and the link strength between the nodes is determined from their cross-correlation (see main text).}
\label{fig1}
\end{center}
\end{figure}

The El~Ni\~no-Southern Oscillation (ENSO) \cite{Wang2017,Timmermann2018, McPhadden2020} is a coupled ocean-atmosphere phenomenon that occurs in the tropical Pacific and can affect global weather and climate. Its three phases are characterized by the sea surface temperature anomaly (SSTA) in the Ni\~no3.4 region (see Fig. \ref{fig1}). An El~Ni\~no is traditionally defined to be present if the 3-month running average SSTA in this region, i.e., the Oceanic Ni\~no Index (ONI)\cite{NOAA}, is above or equal to +0.5°C for at least five consecutive months. La Ni\~na is defined analogously, but the ONI has to be below or equal to -0.5°C. If neither condition is fulfilled, then ENSO is considered to be in a neutral state.

El~Ni\~no and La~Ni\~na episodes can alter precipitation and temperature patterns over a large part of the globe, resulting in, depending on the ENSO phase and region, e.g., flooding or droughts \cite{Wen2002,Corral10,Donnelly07,Kovats03,Davis2001,McPhadden2020}. Early-warning methods are therefore highly desirable to extend the available time to prepare and implement mitigation measures. In general, there are two types of ENSO forecasting approaches: (i) dynamical coupled general circulation models (GCMs), which directly simulate the evolution of the system based on the governing physical laws, for instance, the laws of fluid dynamics, and (ii) statistical models, which rely on statistical relationships between current properties/state of the system and the system’s consequent evolution. These statistical relationships are derived from past data, and the concrete underlying physical processes may often not be known. Numerous models of both types have been proposed to forecast the pertinent index
\cite{Cane86,Penland1995,Tziperman97,Kirtman03,Fedorov03,Galanti03,Chen04,Palmer2004,Luo08,Chen08,Chekroun11,Saha2014, Chapman2015, Lu2016, Feng2016,Rodriguez2016, Nootboom2018, Meng2018, Ham2019,DeCastro2020,Petersik2020,Hassanibesheli2022,Zhao2024,Zhao2024b,Schloer2024}.

Current operational forecasts rely on both types of forecasting approaches, as well as expert judgement, however, they have quite limited anticipation power. In particular, they generally fail to overcome the so-called ``spring predictability barrier'' (see, e.g., \cite{Webster1995,Goddard2001,McPhadden2020}), which usually shortens their reliable warning time to around 6 months \cite{Barnston2012, McPhadden2020, Ehsan2024}. During boreal spring, even confident multi-model ensemble forecasts, i.e., forecasts where more than 75\% of the ensemble members concur, often lead to El~Ni\~no false alarms \cite{Levine2025}.

To resolve this problem, we have recently introduced two alternative forecasting methods, which considerably extend the probabilistic prediction horizon. The first method \cite{Ludescher2013, Ludescher2014, Bunde2024} is based on complex network analysis \cite{Tsonis2006,Yamasaki2008,Donges2009,Gozolchiani2011,Lu2018,Dijkstra2019,Fan2020,Ludescher2021,Fan2022,Lu2022} and provides forecasts for the onset of an El~Ni\~no or its absence in the calendar year before the event starts. The second method \cite{Meng2019} is based on an information entropy, the System Sample Entropy (SysSampEn), which measures the complexity (disorder) in the Ni\~no3.4 area. This approach provides forecasts for both the onset and magnitude of an El Ni\~no event at the end of the previous calendar year. By considering additional predictors, the El~Ni\~no forecasts of these two methods can be further leveraged to obtain more specific forecasts, e.g., for the type (Eastern Pacific or Central Pacific) of an El~Ni\~no event \cite{Ludescher2022,Ludescher2023b}.

For instance, the last real-time El~Ni\~no forecast \cite{Ludescher2023a} based on these methods turned out to be correct. Based on data until December 2022, both methods forecasted the onset of an El~Ni\~no, with a combined probability of around 89\%. The complexity-based approach forecasted a magnitude of $1.49\pm0.37$°C, and the event magnitude was 2.0°C \cite{NOAA}, which is only 1.38 standard deviations from the mean estimate. We also forecasted with an 87.5\% probability that an El~Ni\~no event starting in 2023 would be an Eastern Pacific event, as it turned out to be.

Here, we present the forecasts of both methods for 2026. The climate network-based method predicts the absence of an El Ni\~no in 2026 with 91.4\% probability, while the SysSampEn entropy predicts with 71.4\% probability a weak El Ni\~no with a magnitude of $0.84\pm0.36$°C. In the hindcasted and forecasted past (1984-2025), there were 6 cases where an El Ni\~no onset prediction by the complexity-based method was not matched by the climate network-based method. In 2 of these cases, an El Ni\~no did start (2004, 2006) and in 4 cases, it did not start (1985, 1993, 2012, 2022). Based on this, it appears more likely that an El Ni\~no will not start in 2026. The two diverging forecasts may be combined using Laplace's rule of succession \cite{Jaynes2003}, yielding a (2+1)/(6+2) = 37.5\% probability of an El Ni\~no onset in 2026.

Based on this, the El Ni\~no probability forecasted by our combined methods is lower than the current official NOAA Climate Prediction Center (CPC) probabilistic forecast issued in January 2026 \cite{CPC}. Here, the forecasted El Ni\~no probability is 61\% for the longest available lead time (the target season August-September-October 2026).

A possible reason for the diverging forecasts between the climate network and the complexity-based methods might be the different reanalysis products (NCEP reanalysis 1 and ERA5) on which the two methods are based. We also applied the climate network-based method on ERA5 data. The results suggest a higher El Ni\~no probability than obtained from the original NCEP reanalysis 1 data, more in line with the ERA5-based SysSampEn; see discussion below. We also applied the method described in \cite{Ludescher2023b} and found that, in the case of an El Ni\~no onset, the event will be of the Eastern Pacific type with 76.9\% probability. Finally, we regard the interannual ONI relationship \cite{Ludescher2025} and find that in the absence of an El Ni\~no, a neutral event is favored.

The manuscript is organized as follows. Sections 2 and 3 summarize, respectively, the climate network and SysSampEn-based methods for forecasting El Ni\~no events, discuss their forecasting skills and provide their forecasts for 2026. Section 4 discusses the forecast for the El Ni\~no type.
Section 5 presents the forecast for the non-El Ni\~no outcomes (La Ni\~na and a neutral event).

\section{Climate network-based forecasting}

\subsection{The network-based forecasting algorithm}

For a brief overview of the climate network-based approach, we follow \cite{Ludescher2023a}. The approach is based on the observation that a large-scale cooperative mode, linking the central and eastern equatorial Pacific with the rest of the tropical Pacific (see Fig. \ref{fig1}), builds up in the calendar year preceding an El~Ni\~no event. According to \cite{Gozolchiani2011,Ludescher2013,Ludescher2014}, a measure of emerging cooperativity can be derived from the time evolution of the teleconnections (``links``) between the surface air temperature anomalies (SATA) at the grid points (''nodes``) between the two areas. The strengths of these links are derived from the respective cross-correlations (for details, e.g.,\cite{Ludescher2013,Ludescher2014}).

\begin{figure}[t!]
\begin{center}
\includegraphics[width=15cm]{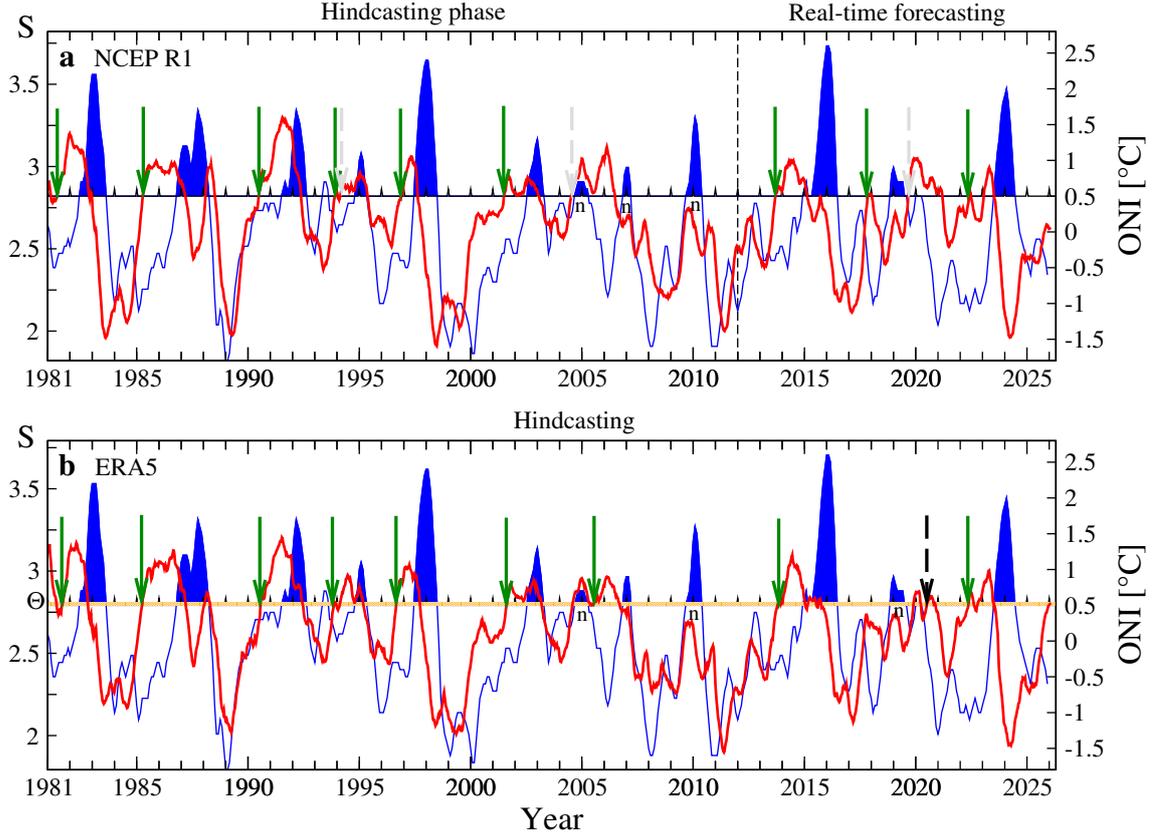}
\caption{{\bf The network-based forecasting scheme.} ({\bf a}) We compare the NCEP reanalysis 1 based  average link strength $S(t)$
in the climate network (red curve) with a decision threshold $\Theta$ (horizontal line, here $\Theta = 2.82$), (left scale), and the Oceanic Ni\~no Index (ONI), (right scale), between January 1981 and December 2025 (present). Please note that we show the 2025 version of the ONI values here. When the link strength crosses the threshold from below, and the last available ONI is below $0.5^\circ$C, we give an alarm and predict that an El~Ni\~no episode will begin in the following calendar year. The El~Ni\~no episodes (when the ONI is at or above $0.5^\circ$C for at least 5 months) are shown by the solid blue areas. Correct predictions are marked with green arrows and false alarms with dashed grey arrows. The index $n$ marks not-predicted events. The threshold was learned in a learning phase between 1950 and 1980 \cite{Ludescher2013} (not shown). The black vertical dashed line separates the hindcasting and real-time forecasting phases. Between 1981 and December 2025, there were 12 El Ni\~no events. The algorithm generated 13 alarms, and 9 of them were correct. In the more restrictive version (ii) of the algorithm, only those alarms are considered where the ONI remains below 0.5°C for the rest of the year. In this version, the incorrect alarms in 1994, 2004, 2019 and 2023 are not activated. Between 1981 and December 2025, version (ii) of the algorithm gave 9 alarms, all of which were correct. In 2025, the average link strength $S(t)$ increased from the very low values in 2024 but remained below the critical threshold band throughout the year, thus forecasting the absence of an El~Ni\~no in 2026. ({\bf b}) Analogous to ({\bf a}), but here $S(t)$ is based on the near-surface air temperature data from ERA5 (1000hPa). We consider version (ii) of the algorithm. Both data sets yield nearly the same forecasts, except for a false alarm in 2020 (ERA5) and correct alarms in 2005 (ERA5) and 2017 (NCEP). By the end of 2025, $S(t)$ based on ERA5 has crossed the lowest threshold, but not all thresholds. This might be interpreted as a partial alarm, potentially consistent with a weak El Ni\~no as forecasted by the SysSampEn (see below).
}
\label{fig2}
\end{center}
\end{figure}

The primary predictive quantity for the onset of an El~Ni\~no is the mean link strength $S(t)$ in the considered network obtained by averaging over all individual links at time $t$ \cite{Ludescher2013,Ludescher2014}. The mean link strength $S(t)$ typically rises in the calendar year before an El~Ni\~no event starts and drops with the onset of the event (see Fig. \ref{fig2}). This property serves as a precursor for the event. The forecasting algorithm involves as only fit parameter, a decision threshold $\Theta$, which has been fixed in a learning phase (1950-1980) \cite{Ludescher2013}. Optimal forecasts in the learning phase are obtained for $\Theta$ between 2.815 and 2.834 \cite{Ludescher2013, Ludescher2014}.

The algorithm issues an alarm and predicts the onset of an El~Ni\~no event the following calendar year when $S(t)$ crosses $\Theta$ from below while the most recent ONI value is below $0.5^\circ$C. In a more restrictive version (ii) \cite{Ludescher2022, Ludescher2023b}, the algorithm considers only those alarms where the ONI remains below 0.5°C for the rest of the calendar year.

For the calculation of $S$, we use daily surface air temperature data from the National Centers for Environmental Prediction/National Center for Atmospheric Research (NCEP/NCAR) Reanalysis 1 project \cite{reanalysis1,reanalysis2}, as the method was originally introduced based on this data set \cite{Ludescher2013, Ludescher2014}. We like to note that the approach has a high predictive skill when based on observational (reanalysis) data (see below and also \cite{Hu2022, Bunde2024}). In contrast, when applied analogously within CMIP5 or CMIP6 historical and control runs, the predictive skill is absent or very low \cite{Ludescher2022c, Han2026}. The lack of this predictive feature may explain the GCMs' difficulties in overcoming the spring barrier and may be related to biased air-sea interactions \cite{Han2026}.

\subsection{El~Ni\~no forecasts since 2011}
The climate network-based algorithm has been quite successful in providing real-time forecasts, i.e., forecasts into the future. In its original version, it provided 12 forecasts for the period 2012-2023; 11 of these forecasts turned out to be correct (see Fig. \ref{fig2}). The only incorrect forecast was a false alarm issued in September 2019. In December 2022 \cite{Ludescher2022, Ludescher2023b}, we introduced a more restrictive version (ii) of the algorithm, where only those alarms are considered where the ONI remains below 0.5°C for the rest of the year, and we have been using this version exclusively from then on. Version (ii) correctly forecasted the absence of an El~Ni\~no onset in 2024. Thus, in total, the method provided 14 real-time forecasts for the period 2012-2025, 13 of these forecasts turned out to be correct. The p-value, obtained from random guessing with the climatological El~Ni\~no onset probability (for details, see \cite{Bunde2024}), for the skill in the forecasting period is $p\cong3.1\cdot10^{-3}$. When considering the hindcasting and forecasting periods (1981-2025) correspondingly together, the p-value is $p\cong1.9\cdot10^{-5}$.

In the hindcasting and forecasting period (January 1981 to December 2025), there were 12 El~Ni\~no events based on the 2025 version of the ONI. Version (ii) of the algorithm \cite{Ludescher2022, Ludescher2023b} gave 9 El~Ni\~no alarms, all of which were correct. Equivalently, it forecasted 35 times the absence of an El~Ni\~no onset and missed 3 events. Thus, the forecasts for the absence of an El~Ni\~no onset are correct with 32/35 $\approx$ 91.4\% probability. In 2025, the average link strength $S(t)$ increased from the very low values in 2024 but remained below the critical threshold band throughout the year, thus forecasting the absence of an El~Ni\~no in 2026.

\subsection{Climate network forecast based on ERA5}
The climate network-based forecasting approach was introduced based on the NCEP reanalysis 1 data set. Also, the real-time forecasts \cite{Ludescher2019, Ludescher2021b, Ludescher2022b, Ludescher2022, Ludescher2023a, Ludescher2024, Ludescher2025} were based on this data set. Since the SysSampEn-based approach, which is based on ERA5, predicts an El Ni\~no (see below), we test the climate network-based approach on ERA5 daily near-surface (1000hPa) temperature data. For consistency with the NCEP reanalysis 1 data, we use the data from 1948 to the present and calculate the anomalies analogously. To obtain a narrow threshold band, and since the general approach has already been validated on NCEP reanalysis 1 and in real-time forecasts, we use the full dataset to learn the optimal thresholds. We use version (ii) of the algorithm and find that the best forecasting performance based on the Heidke skill score (HSS) is obtained for $2.792\leq\Theta\leq2.810$. For these thresholds, we obtain 13 correct hindcasts and 3 false alarms over the full period (1950-2025). Figure \ref{fig2}b shows the forecasts for the period 1981 - present. The range of the critical thresholds is shown as a narrow orange band. By the end of 2025, $S(t)$ increased up to 2.793, thus crossing the lowest threshold, however, not all thresholds have been crossed. Also, the threshold band is typically clearly crossed well before the end of a calendar year (e.g., 1996), and by the end of a year, $S(t)$ is well above the 2025 value.

Liberally interpreted, the crossing of the lowest threshold could be equated with a full crossing, thus creating an El Ni\~no onset alarm. Based on the hindcasting performance, this interpretation corresponds to a 13/16 = 81.25\% probability of an El Ni\~no onset. Assuming this is the correct interpretation, an equally weighted average of the NCEP- and ERA5-based El Ni\~no probabilities yields a combined probability of (8.6\% + 81.25\%)/2 = 44.9\% that an El Ni\~no will start in 2026. Since this value represents an upper bound, it aligns well with the combined probability discussed in the Introduction. Please note that this probability can not simply be combined with the SysSampEn-based  forecast, as it would double-count the ERA5-based forecasts.

Most strictly interpreted, only the crossing of the highest threshold creates an El Ni\~no onset alarm. Such an interpretation would predict the absence of an El Ni\~no in 2026 with $48/59 \approx 81.4\%$ probability, since based on ERA5, 11 El Ni\~nos were missed in total, while 59 El Ni\~no absences were hindcasted. A detailed probability quantification of this partial crossing, which is uncommon in practice, will be addressed in future work.

\section{System Sample Entropy-based forecast}

\subsection{SysSampEn} 

\begin{figure}[]
\begin{center}
\includegraphics[width=9.2cm]{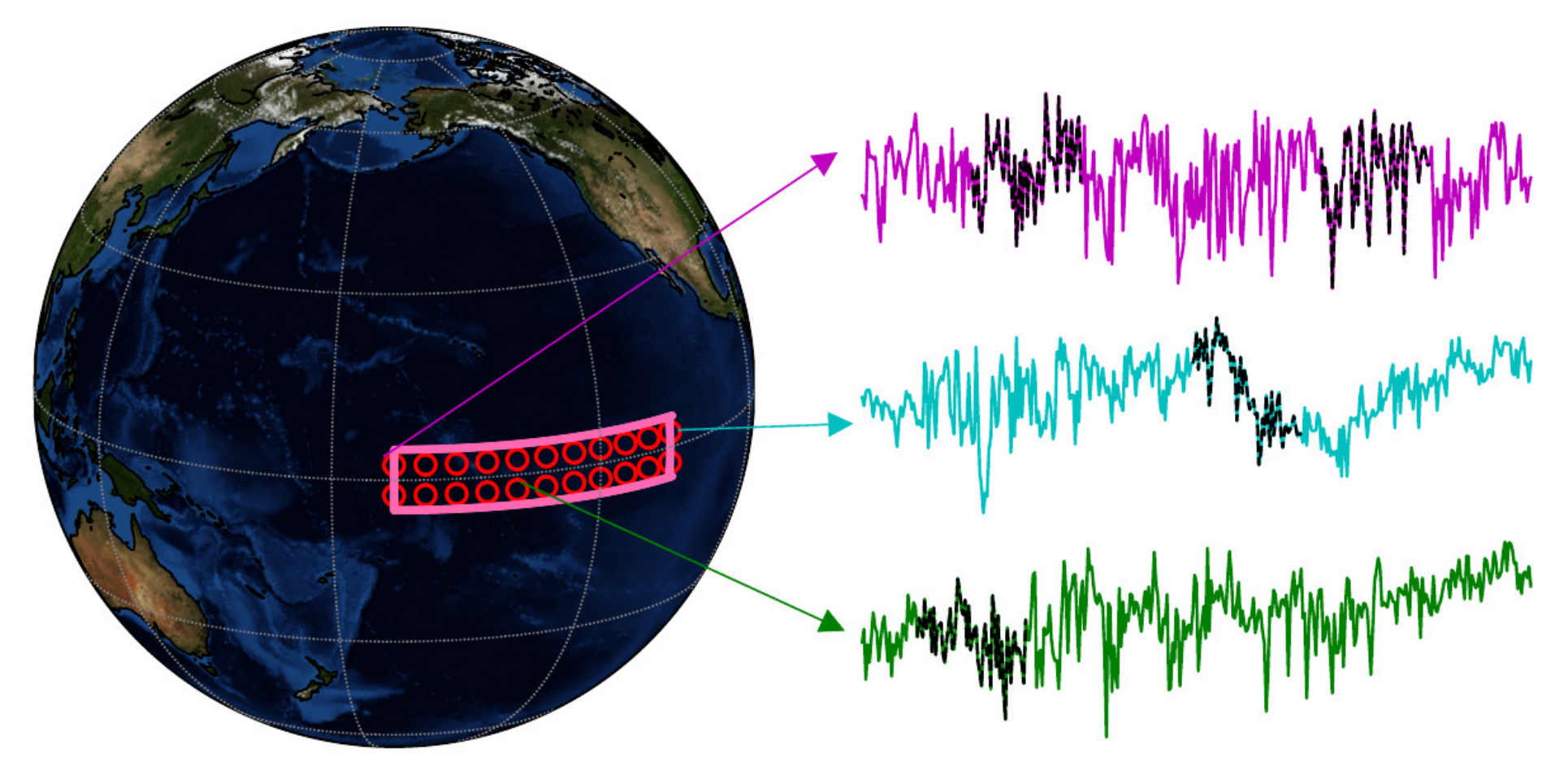}
\caption{{\bf The Ni\~no3.4 area and the SysSampEn input data.}
The red circles indicate the 22 nodes covering the Ni\~no 3.4 region at a spatial resolution of $5^\circ \times 5^\circ$. The curves are examples of the temperature anomaly time series for 3 nodes in the Ni\~no 3.4 region for one specific year. Several examples of their subsequences are marked in black. In the calculation of the SysSampEn, both the similarity of subsequences within a time series and that of subsequences across different time series are considered.
Figure from \cite{Meng2019}.}
\label{fig3}
\end{center}
\end{figure}

For a brief description of the System Sample Entropy (SysSampEn) approach, we follow \cite{Ludescher2023a}. The SysSampEn was introduced in \cite{Meng2019} as an analysis tool to quantify the complexity (disorder) in a complex system, in particular, in the temperature anomaly time series in the Ni\~no3.4 region. The SysSampEn is approximately equal to the negative natural logarithm of the conditional probability that 2 subsequences similar (within a certain tolerance range) for $m$ consecutive data points remain similar for the next $p$ points, where the subsequences can originate from either the same or different time series (e.g., black curves in Fig. \ref{fig3}), that is,
\begin{equation}
SysSampEn(m, p, l_{eff}, \gamma) = -log(\frac{A}{B}),
\end{equation}
where A is the number of pairs of similar subsequences of length $m + p$, $B$ is the number of pairs of similar subsequences of length $m$, $l_{eff} \leq l$ is the number of data points used in the calculation for each time series of length $l$, and $\gamma$ is a constant that determines the tolerance range. The definition of the SysSampEn for a general complex system composed of $N$ time series and how to objectively choose the parameter values are described in detail in \cite{Meng2019}.

In \cite{Meng2019}, it was found that the previous year's ($y-1$) SysSampEn exhibits a strong positive correlation ($r=0.90$ on average) with the magnitude of an El Ni\~no in year $y$ when parameter combinations are used that are able to quantify a system's complexity with a high accuracy. This linear relationship between SysSampEn and El Ni\~no magnitude thus enables predicting the magnitude of an upcoming El Ni\~no when the current ($y-1$) SysSampEn is inserted into the linear regression equation between the two quantities.

If the forecasted El~Ni\~no magnitude is below $0.5^\circ$C then the absence of an El Ni\~no onset is predicted for the following year $y$. Thus, SysSampEn values below a certain threshold indicate the absence of an El Ni\~no onset. In contrast, if the SysSampEn is above this threshold and the ONI in December of the current year is below $0.5^\circ$C, then the method predicts the onset of an El~Ni\~no event in the following year.

\subsection{Forecast for 2026}

\begin{figure}[t]
\begin{center}
\includegraphics[width=15cm]{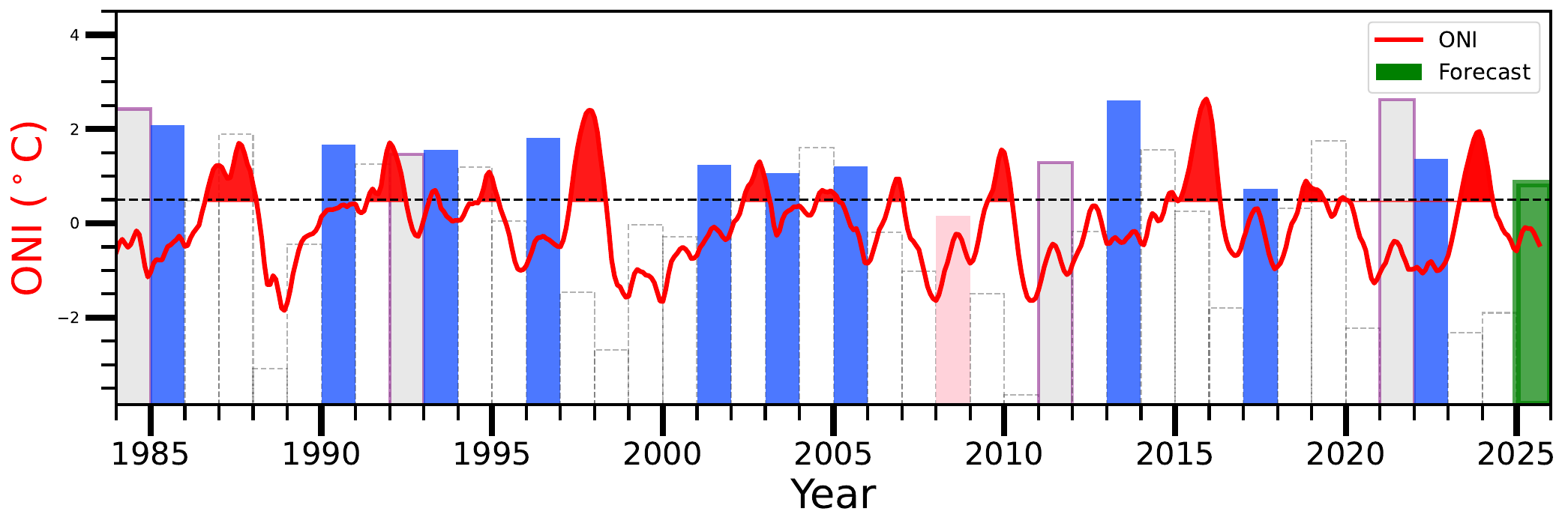}
\caption{{\bf Forecasted and observed El~Ni\~no magnitudes.} The magnitude forecast is shown as the height of rectangles in the year when the forecast is made, i.e., one year ahead of a potential El~Ni\~no. The forecast is obtained by inserting the regarded calendar year's SysSampEn value into the linear regression function between SysSampEn and El~Ni\~no magnitude. To forecast the following year's condition, we use the ERA5 daily near-surface (1000 hPa) temperatures with the set of SysSampEn parameters ($m = 30$, $p = 30$, $\gamma=8$ and $l_{eff} = 360$). The red curve shows the 2025 version of the ONI, and the red shades highlight the El~Ni\~no periods. The blue rectangles show the correct prediction of an El~Ni\~no in the following calendar year. The onset of an El~Ni\~no in the following year is predicted if the forecasted magnitude is above $0.5^\circ$C and the current year's December ONI is $<0.5^\circ$C. White rectangles show correct forecasts for the absence of an El~Ni\~no. Grey bars with a violet border show false alarms, and the only missed event is shown as a pink rectangle. The SysSampEn value for 2025 is $1.60$, which is above the threshold value of 1.50. There were 14 occurrences of high SysSampEn accompanied by a low ONI in December, as is the case in 2025 (green rectangle). In 10 out of these 14 cases, the hindcast was correct. Thus, the method predicts with 71.4\% probability the onset of an El~Ni\~no in 2026.
}
\label{fig4}
\end{center}
\end{figure}

Here we use as input data the daily near-surface (1000 hPa) air temperatures from the ERA5 reanalysis of  the European Centre for Medium-Range Weather Forecasts (ECMWF) \cite{ERA5} analysed at a $5^\circ$  \ resolution. The last months of 2025 are based on the initial data release ERA5T, which, in contrast to ERA5, lags only a few days behind real-time.

The daily time series are preprocessed by subtracting the corresponding climatological mean and then dividing by the climatological standard deviation. We start in 1984 and use the previous years to calculate the first anomalies. For the calculation of the climatological mean and standard deviation, only past data up to the year of the prediction are used. For simplicity, leap days are excluded. We use the parameter values for the SysSampEn that lead to the best El~Ni\~no forecasting skill when applied to past events, as described in \cite{Meng2019}, $m = 30$, $p = 30$, $\gamma=8$ and $l_{eff} = 360$.

Figure \ref{fig4} shows the analysis results. The magnitude forecast is shown as the height of rectangles in the year when the forecast is made, i.e., 1 year ahead of a potential El~Ni\~no onset. The forecast is obtained by inserting the regarded calendar year's SysSampEn value into the linear regression function between SysSampEn and El~Ni\~no magnitude. For the 2026 forecast, the regression is based on all correctly hindcasted El~Ni\~no events before 2025. The red curve shows the ONI and the red shades indicate the El~Ni\~no periods. The blue rectangles show the correct prediction of an El~Ni\~no in the following calendar year and grey rectangles with a violet border show false alarms. White rectangles show correct forecasts for the absence of an El~Ni\~no, and the only missed event is the 2009/10 El~Ni\~no, for which the preceding SysSampEn value was slightly below the threshold.

There were 14 occurrences of a high SysSampEn accompanied by an ONI below 0.5°C in December. In 10 out of these 14 cases, the hindcast was correct. The forecasted magnitude of an El Ni\~no in 2026 is $0.84\pm0.36$°C, as shown by the green rectangle. The SysSampEn value for 2025 is 1.60, which is above the threshold value of 1.50. Thus, the method predicts with 71.4\% probability the onset of an El~Ni\~no in 2026.

\section{El Ni\~no type forecast}

El Ni\~no episodes can exhibit their maximal warming in the equatorial Central Pacific (CP) or Eastern Pacific (EP). The type of an El Ni\~no (CP or EP) also has a major influence on the event's impact, which can even lead to either dry or wet conditions in the same region. In \cite{Ludescher2023b}, we proposed an index, $\Delta T_{WP-CP}$, for forecasting the type of an El Ni\~no by the end of the calendar year before the event starts. The index is based on the difference of the sea surface temperature anomalies between the equatorial western and central Pacific. Analogously, the index can also be defined using sea surface height (SSH) or surface air temperature (SAT) anomalies and leads to nearly the same forecasts \cite{Ludescher2023b}. When the index is positive, an El Ni\~no starting in the following calendar year will likely be an EP event, otherwise a CP event. Interestingly, all CP hindcasts are correct, whereas only 10 out of 13 EP hindcasts and forecasts turned out correct \cite{Ludescher2023b}. The difference in correct forecasts may be explained by westerly wind events (WWE) and ocean preconditioning, for a discussion, see \cite{Ludescher2023b}. Please note that the index, by itself, does not forecast the onset of an El Ni\~no or its absence. It is thus most useful when combined with an approach that can forecast the onset of an El Ni\~no event, such as the climate network and SysSampEn-based methods, but also coupled general circulation models (GCMs).

In December 2025, $\Delta T_{WP-CP}=+0.76$°C when based on the Extended Reconstructed Sea Surface Temperature version 5 (ERSSTv5)\cite{Huang2017} data, which indicates an EP type event. A positive value is also confirmed by SSH and SAT data. Thus, if an El Ni\~no starts in 2026, with $10/13 \approx 76.9\%$ probability, it will be an EP event. We like to note that since EP events tend to be stronger than CP events, an EP type forecast would be more in line with a larger El Ni\~no magnitude than the weak El Ni\~no forecasted by the SysSampEn. Nevertheless, there were a few weak EP El Ni\~nos in the past, e.g., the 2006/07 event.

\section{Probability of a La Ni\~na vs. a neutral event}

The above approaches provide forecasts focused on the El Ni\~no phase of ENSO. In \cite{Ludescher2025}, we proposed the interannual relationship of ONI values as an additional predictor to further specify the forecast and discriminate between a La Ni\~na and a neutral event. The predictor is based on the observation that consecutive end-of-year ONI values exhibit some structure, that is, they do not follow each other randomly. For instance, in the observational record since 1950, neutral events were not followed by La Ni\~nas. Based on this method combined with our El Ni\~no forecasting methods, we forecasted in real-time that boreal winter 2025/2026 would be a neutral event with a 69.6\% probability. At the time of writing, it appears very likely that the forecast will turn out to be correct. Please note that throughout the manuscript, we refer to the traditional ONI-based classifications rather than the relative ONI (RONI)\cite{Oldenborgh2021, LHeureux2024}, which NOAA adopted after our forecast.

\begin{figure}[h]
\begin{center}
\includegraphics[width=9cm]{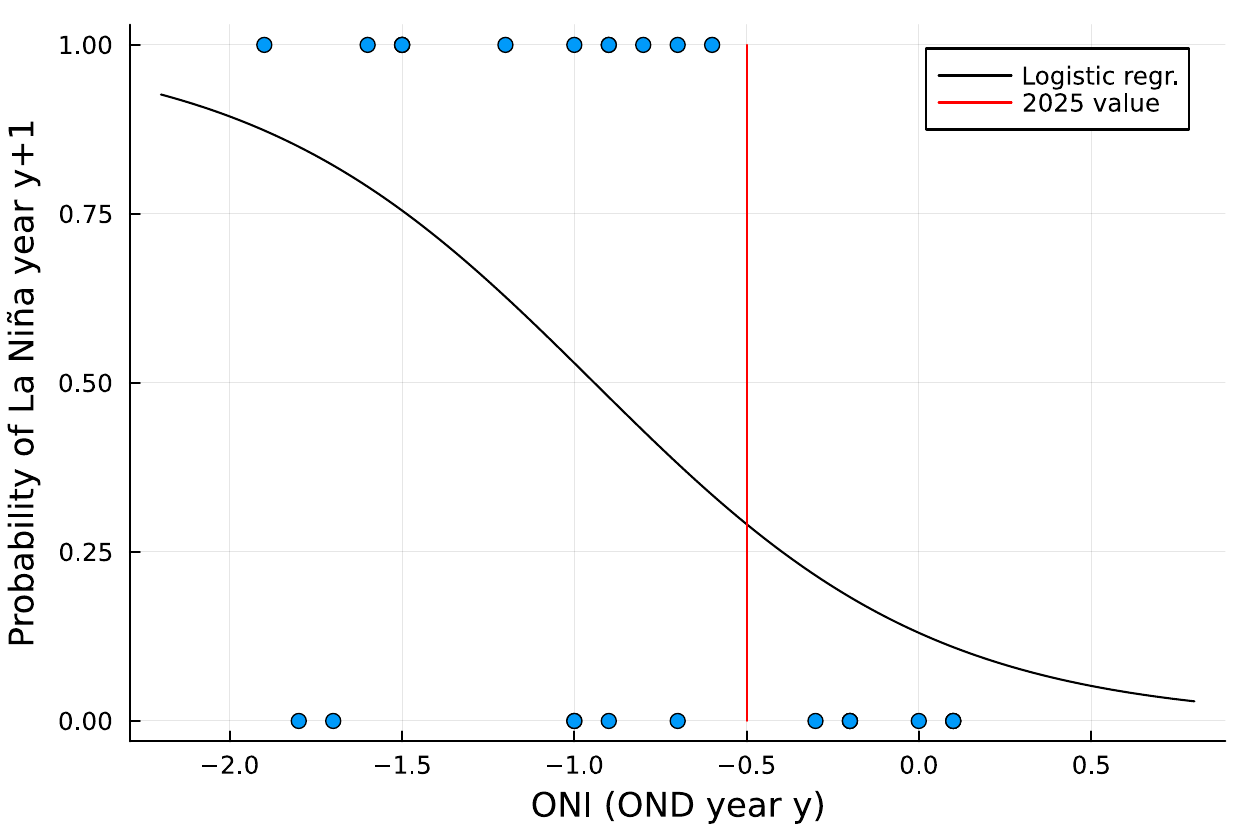}
\caption{{\bf The probability of a La~Ni\~na}. We regard all years that are non-El~Ni\~no years and are also followed by a non-El~Ni\~no year. Under the assumption that no El Ni\~no starts in 2026, these events correspond to the current state. The outcome of the second year is encoded as 1 for La~Ni\~na and 0 for a neutral event (blue circles). To obtain the probability of a La~Ni\~na event in 2026, given that no El~Ni\~no starts in 2026, we apply a logistic regression and use the current OND ONI value of $-0.5$°C as a predictor. Excluding El~Ni\~no, we obtain a $29.1\%$ probability for a La~Ni\~na vs. a $70.9\%$ probability for a neutral event.
}
\label{fig6}
\end{center}
\end{figure}

The current October-November-December (OND) ONI value is -0.5°C, i.e., we are clearly not in an El Ni\~no phase. To obtain the probability of a La Ni\~na vs. a neutral event, we regard all non-El Ni\~no (i.e., La Ni\~na or neutral) years that are also followed by a non-El Ni\~no year \cite{Ludescher2025}. Figure \ref{fig6} shows the OND ONI values of these events in year $y$ and the outcome in year $y+1$. La Ni\~nas are encoded as 1 and neutral events as 0. The probability is obtained via a logistic regression. Based on the current OND ONI value of -0.5°C, we obtain 29.1\% and 70.9\% probabilities for a La Ni\~na and a neutral event, respectively, given that no El Ni\~no starts in 2026. To obtain the unconditioned probability, these values have to be multiplied by the complementary El Ni\~no onset probability ($1-p_{El Nino}$). Assuming the El Ni\~no onset probability is 37.5\% (see Introduction), this yields a $(1-0.375)\times0.291 = 18.2\%$ probability for a La Ni\~na and correspondingly a 44.3\% probability for a neutral event.

\begin{figure}[]
\begin{center}
\includegraphics[width=11cm]{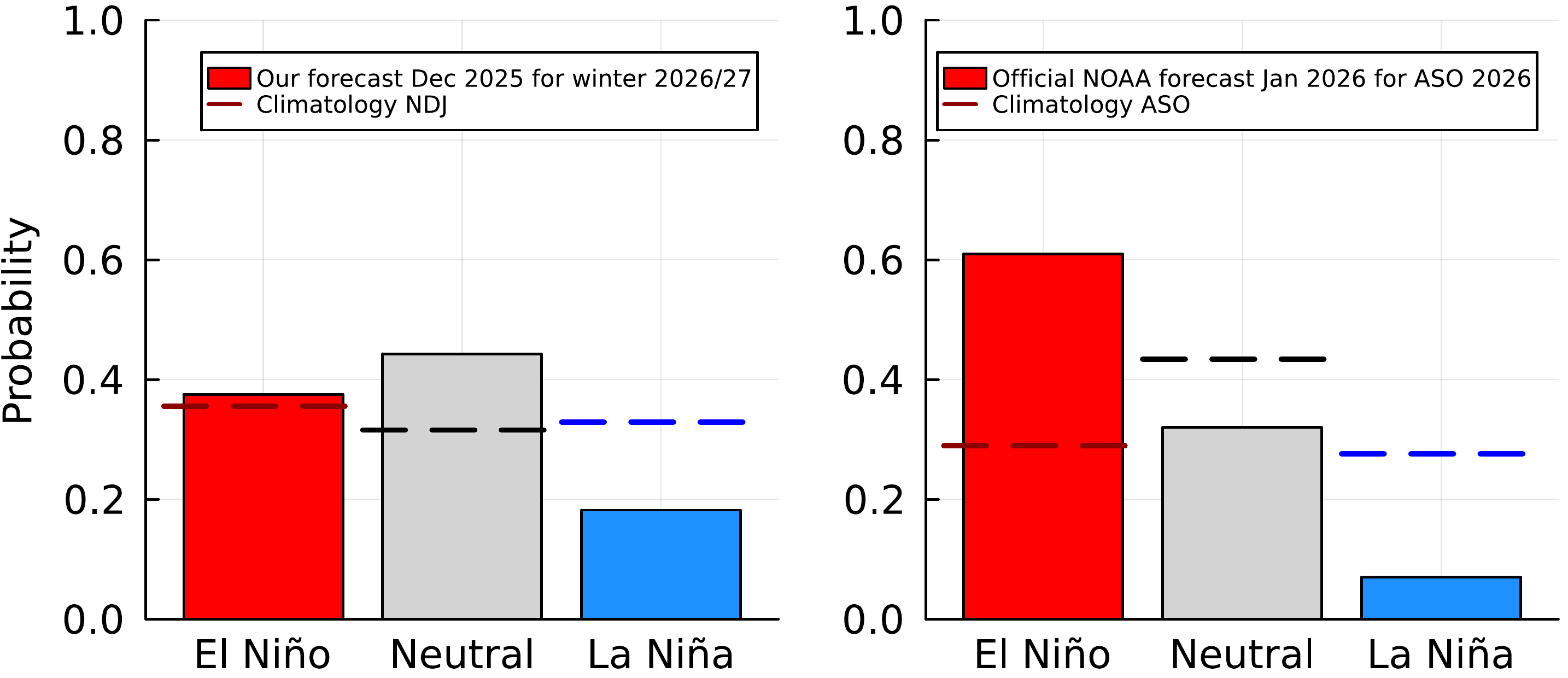}
\caption{{\bf Summary of our forecast.} The height of the bars shows the probability of El~Ni\~no, neutral event and La~Ni\~na for NDJ 2026/27 based on the climate network, the SysSampEn and the interannual relationship of ONI values. We obtain $37.5\%$, $44.3\%$ and $18.2\%$, respectively. The corresponding probabilities of the NOAA CPC forecast from January 2026 are $61\%$, $32\%$ and $7\%$ for the target season ASO 2026 \cite{CPC}. The dashed horizontal lines show the NDJ and ASO climatological probabilities, respectively.
}
\label{fig7}
\end{center}
\end{figure}

Figure \ref{fig7} summarizes our final forecast probabilities for boreal winter 2026/27, i.e., the probabilities that NDJ 2026/27 will be part of the respective ENSO phases. We compare these probabilities with the climatologically expected outcome in NDJ (dashed horizontal lines). Figure \ref{fig7} also shows the corresponding official probabilistic forecast of the NOAA Climate Prediction Center (CPC) for August-September-October (ASO), which was issued in January 2026 \cite{CPC}. We show ASO since this is the longest lead time prediction target provided by this forecast. Compared with the NOAA CPC forecast, our combined forecast indicates a lower probability of an El Ni\~no event in 2026, with the estimate only slightly exceeding the climatological baseline.

\newpage

\section*{Acknowledgements}
J. L. is part of the Planetary Boundaries Science Lab's research effort at PIK. J. M. was supported by the National Natural Science Foundation of China (Grant No. 42575057, T2525011, 42450183, 12275020, 42461144209). J. F. was supported by the National Key R\&D Program of China (2025YFF0517203, 2025YFF0517304).

\section*{Data Availability}
The National Centers for Environmental Prediction/National Center for Atmospheric Research (NCEP/NCAR) Reanalysis 1
surface air temperature data are publicly available from: \url{https://psl.noaa.gov/data/gridded/data.ncep.reanalysis.html}.
The Oceanic Ni\~no Index (ONI) is available at: \url{https://www.cpc.ncep.noaa.gov/products/analysis_monitoring/ensostuff/ONI_v5.php}. The European Center for Medium-Range Weather Forecast (ECMWF) Reanalysis v5 (ERA5) near-surface air temperature data are available from: \url{https://climate.copernicus.eu/climate-reanalysis?q=products/
climate-reanalysis}. The Extended Reconstructed Sea Surface Temperature (ERSST) v5 data is available at \url{https://www.ncei.noaa.gov/
products/extended-reconstructed-sst}.

\end{document}